# Low temperature Spin freezing and Diffuse Magnetic Correlations in $Tb_2Zr_{2-x}Ti_xO_7$ ($x$ = 0, 0.5)

Sujata Singh[1], Leon Carstens[2], M. Duc Le[3], R. Klingeler[2], C S Yadav[1,4*]

[1]School of Physical Sciences, Indian Institute of Technology Mandi, Mandi-175005 (H.P.) India
[2]Kirchhoff Institute of Physics, Heidelberg University, INF 227, D-69120 Heidelberg, Germany
[3]ISIS Neutron and Muon Source, STFC Rutherford Appleton Laboratory, Didcot OX11 0QX, UK
[4]Center for Quantum Science and Technologies, Indian Institute of Technology Mandi, Mandi-175005 (H.P.) India

Structural disorder in the magnetically frustrated pyrochlore system leads to intriguing magnetic states. We present the thermodynamic behavior and short-range magnetic correlations in the $Tb_2Zr_2O_7$ and $Tb_2Zr_{1.5}Ti_{0.5}O_7$ compounds. The parent compound $Tb_2Zr_2O_7$ has defect fluorite structure, which evolves toward the pyrochlore phase on $Ti$ doping at $Zr$ site. There is no long-range magnetic order down to 0.4 K, and a magnetic field dependent spin freezing evolves below ~ 1.25 K and ~ 1.05 K for the parent and doped compounds respectively. The ac susceptibility measurements indicate slow spin relaxation process below ~ 20 K in these compounds. Inelastic neutron scattering reveals broad diffuse scattering, indicative of short-range correlations at low temperature, owing to local structural distortions and persistent spin fluctuations. These results suggest a correlated, disorder-influenced magnetic state in $Tb_2Zr_2O_7$, $Tb_2Zr_{1.5}Ti_{0.5}O_7$ compounds.

## INTRODUCTION

The study of geometrically frustrated pyrochlore materials has revealed a wide spectrum of unconventional magnetic phases and diverse ground states owing to complex magnetic interactions in these lattices of corner-sharing tetrahedra [1]. The magnetic ground state of these systems is immensely influenced by the choice of rare-earth *(R)* and the transition metal *(M)* ions in $R_2M_2O_7$ framework, which dictate the anisotropy and exchange coupling mechanisms [2,3]. The unique lattice geometry frustrates conventional magnetic ordering and, together with anisotropic interactions, promotes the emergence of complex, often disordered, ground states. The intricate interdependence between lattice topology and magnetic interactions continues to enrich the understanding of frustrated magnetism and the development of novel quantum phases.

In the classical spin-ice compounds $Dy_2Ti_2O_7$ and $Ho_2Ti_2O_7$, dominant ferromagnetic Ising interactions enforce the local "two-in, two-out" ice rule on each tetrahedron of the pyrochlore lattice [4, 5]. This leads to a degenerate ground-state, described by emergent magnetic monopoles and long-range Coulombic spin correlations [6, 7]. Geometrical frustration suppresses long-range order, slowing spin dynamics at low Temperature (*T*) as monopoles proliferate and interact, yielding glass-like freezing with relaxation times diverging on experimental scales [7, 8]. In another frustrated pyrochlore, $Tb_2Ti_2O_7$ a correlated, fluctuation-dominated regime persists down to the lowest temperatures, and the competition between crystal-field anisotropy and strong magnetoelastic coupling stabilizes an unusual quantum-disordered state and a proximate spin-liquid like behavior [9, 10]. Further, the chemical disorder at the transition metal ion site or the random site occupancy breaks translational symmetry, and promotes competing quantum correlations at low temperatures [11]. For example, in $Gd_2(Ti_{1-x}Zr_x)_2O_7$, substitution of $Zr$ for $Ti$ induces an order–disorder transition from the pyrochlore to the defect-fluorite phase through progressive cation anti-site defects and oxygen anti-Frenkel defects [12]. The pyrochlore phase *(Fd–3m)* features distinct Wyckoff positions of *R* and *M* cations with minimal structural disorder, whereas the defect-fluorite phase *(Fm–3m)* possesses random distributions of *R* and *M* cations at *4a* position along with the oxygen-vacancy disorder. The $Tb_2Zr_2O_7$ crystallizes in the defect-fluorite structure and exhibits a collective, unconventional magnetic response with field-induced spin relaxation, similar to $Tb_2Ti_2O_7$ [9,10,13].

In this work, the low-*T* (down to 0.4 K) magnetization and crystal-field scheme of $Tb_2Zr_2O_7$ and *Ti*-substituted $Tb_2Zr_{1.5}Ti_{0.5}O_7$ are investigated, revealing the onset of spin freezing, its evolution with *M*-site substitution, and the associated magnetic excitations. Low-*T* magnetization measurements reveal the absence of long-range magnetic order, and the emergence of a spin-frozen state below 1.3 K, accompanied by slow spin relaxation that shifts toward higher temperatures (~20 K at 20 kOe) under applied magnetic fields. Diffuse magnetic scattering detected by inelastic neutron scattering measurements indicates short-range correlations without discernible crystal-field excitations, while both compositions display thermal irreversibility between zero-field-cooled and field-cooled curves. Complementary point-charge model calculations incorporating both structural configurations are also discussed within an effective ligand-field model framework to account for the influence of local symmetry on the crystal-field scheme.

**EXPERIMENTS**

Polycrystalline samples of $Tb_2Zr_{2-x}Ti_xO_7$ *(x = 0* and *0.5)* were synthesized via a solid-state reaction route. The $Tb_2O_3$ (99.99%, Sigma Aldrich) was preheated to remove moisture before weighing, then thoroughly mixed with high-purity $ZrO_2$ and $TiO_2$ (both 99.99%) in stoichiometric ratio. The homogenized mixtures were ground for several hours and kept at 1400 °C for 50 hours in alumina crucible [14]. X-ray diffraction (*XRD*) study for phase identification was conducted using a Rigaku D/Max-B benchtop diffractometer equipped with *Cu-K$_α$* radiation ($\lambda$ = 1.541 Å) at room *T* over a range of 10° to 90°. Magnetization measurements in the *T* range 1.8–300 K were carried out using a Quantum Design built SQUID magnetometer. The Reitveld refinement of the XRD pattern was done using FULLPROF SUITE Software package [15]. The crystal structure was analyzed using VESTA software [16]. The dc magnetization measurements were performed over a *T* range of 1.8 ≤ T ≤ 350 K using a Quantum Design Magnetic Properties Measurement System (MPMS3, Quantum Design). For measurements extending down to 400 mK, the MPMS3 was fitted with the iQuantum He3 option. Magnetization data were collected under zero-field-cooled (ZFC) and field-cooled (FC) protocols. AC magnetization measurements were carried out from 1.8 to 50 K, employing 2–8 Oe AC drive fields, DC fields up to 5 T, and frequencies in the range $1 \leq f\,(Hz) \leq 1000$ using the AC module of the MPMS3.

The Inelastic Neutron Scattering (INS) measurements were carried out on the MARI time-of-flight spectrometer at the ISIS Neutron and Muon Source, Rutherford Appleton Laboratory [17]. The experiment was conducted using 4.0 g of the parent powder sample and 4.4 g of the doped powder sample. Each polycrystalline specimen was wrapped in aluminum foil, arranged in an annular geometry, and sealed in a cylindrical aluminum can of 4 cm diameter. The samples were mounted in the instrument's closed-cycle refrigerator (CCR) and a Fermi chopper with the Gd-slit package. We collected data at 4, 10, and 100 K with incident neutron energies of 8.75, 11.2, 27.5, 65, and 150 meV. Normalization to the aluminum standard allowed determination of the dynamic structure factor $S(Q, E)$ on an absolute intensity scale using the MANTID software package [18].

**RESULTS**

A. Crystal structure: Doping induced structural transition

The Rietveld refinement of the XRD data on $Tb_2Zr_{2-x}Ti_xO_7$ (x = 0 and 0.5) compounds showed single-phase XRD patterns (Fig. 1(a)) without any detectable impurities. The cubic fluorite model (space group *Fm–3m*) fits the parent compound well, while the Ti-doped sample is better described by the pyrochlore phase (space group *Fd–3m*). With Ti substitution (x = 0.50), the superlattice reflections at 2θ = 14° (111), 27° (311), 37° (331) and 45° (511), corresponding to the pyrochlore phase emerge clearly. The refined oxygen anion occupancy of 0.815(4) is close to the expected stoichiometric value of 7/8. A stable pyrochlore structure is formed for the compounds having their cation radius ratio ($r_A/r_B$), in the range of 1.46-1.78, and the lower ratio favors a fluorite phase [19]. In the present study, the doped compound shows $r_A/r_B$ is 1.5, which falls within the stability window, making it more structurally stable compared to the parent compound with a ($r_A/r_B$) ratio of 1.4. The Raman spectroscopic data shown in Fig 1(b) also confirm the phase evolution with Ti doping. For $Tb_2Zr_2O_7$, the spectrum displays a broad hump and a single prominent $F_{2g}$ mode, a characteristic of the defect fluorite phase [20]. The broad hump (~200-550 cm$^{-1}$) represents the low-symmetry or disordered phonon modes resulting from oxygen sublattice disorder and such a broad hump is typically found in the fluorite structure [20]. For $Tb_2Zr_{2-x}Ti_xO_7$, distinct signatures of pyrochlore structure emerge, revealing five active modes ($3F_{2g}+A_{1g}+E_g$) associated with the *48f* and *8a* oxygen sites [21]. The observed phonon features include modes at 303 cm$^{-1}$ ($F_{2g}$), corresponding to *Tb-O′* bending vibrations; 364 cm$^{-1}$ ($E_g$), arising from the deformation of the $Tb_4O$ tetrahedral units; 528 cm$^{-1}$ ($A_{1g}$), corresponding to symmetric stretching of *Tb-O + Tb-O′*; 580 cm$^{-1}$ ($F_{2g}$), linked to asymmetric $Tb_4O$ distortions; and 642 cm$^{-1}$ ($F_{2g}$), associated with stretching modes coupled to *Ti-O′* vibrations. Weak additional features near 420 cm$^{-1}$ can be attributed to a minor $TiO_2$ in $Tb_2Zr_{1.50}Ti_{0.50}O_7$. The enhanced intensities of the $F_{2g}$, $A_{1g}$, and $E_g$ modes in the Ti-rich pyrochlore versus Zr in Ti-enriched compositions likely originate from lattice distortions induced by cation radius mismatch, which enhances phonon anharmonicity [22]. The crystal structures of both compounds are depicted in Fig. 1(c) and Fig. 1(d).

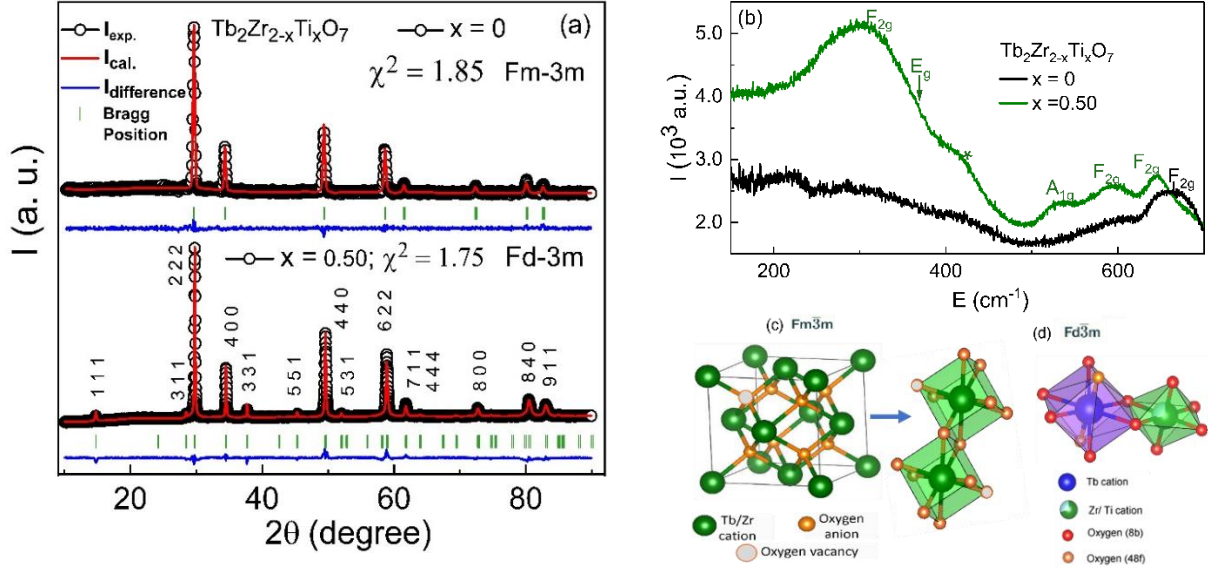

Figure 1. (a) X-ray diffraction (XRD) patterns of the synthesized compounds, confirming phase purity and structural integrity. (b) Room-$T$ Raman spectra revealing vibrational modes characteristic of the respective crystal structures. (c, d) Schematic representations of the fluorite and pyrochlore lattices, highlighting the tetrahedral coordination in the pyrochlore structure.

Table I. Structural parameters obtained from the Reitveld refinement of XRD data

| Atom | Site | x | y | z |
|---|---|---|---|---|
| | $Tb_2Zr_2O_7$ | Space group: Fm-3m | a = 5.223(4) Å | |
| Tb | 4a | 0.000 | 0.000 | 0.000 |
| Zr | 4a | 0.000 | 0.000 | 0.000 |
| O | 8c | 0.250 | 0.250 | 0.250 |
| | $Tb_2Zr_{1.5}Ti_{0.5}O_7$ | Space group: Fd-3m | a = 10.398(9) Å | |
| Tb | 16d | 0.500 | 0.500 | 0.500 |
| Zr | 16c | 0.000 | 0.000 | 0.000 |
| Ti | 16c | 0.000 | 0.000 | 0.000 |
| O | 8b | 0.375 | 0.375 | 0.375 |
| O′ | 48f | 0.331(7) | 0.125 | 0.125 |

B. Signatures of slow spin dynamics

Fig. 2 shows the $T$ dependence of dc magnetic susceptibility ($\chi(T) = M/H$) measured at an applied field of H = 100 Oe over the $T$ range 1.8 K ≤ T ≤ 350 K. Both compounds exhibit paramagnetic behavior with no bifurcation between zero-field-cooled (ZFC) and field-cooled (FC) protocols [insets of Fig. 2(a, b)]. This indicates the absence of long-range magnetic order or phase transitions, is consistent with observations in both dc susceptibility and frequency-dependent zero-field AC susceptibility measurements. The $\chi(T)$ data above 50 K follow the free ion Curie-Weiss law, $\chi(T) = C/(T - \theta_C)$, where C is Curie constant and $\theta_C$ is Curie-Weiss $T$. The fitting in the range 50 ≤ T ≤ 350 K (red lines in Fig. 2(a, b)) yield negative $\theta_C$ values, indicating dominant AFM interactions among $Tb^{3+}$ ions. The obtained effective moments ($\mu_{eff}^{HT} =$

$\sqrt{8C}$ $\mu_B$) are 9.54 and 9.59 $\mu_B/Tb^{3+}$ for the parent and doped compounds, respectively, that is lightly lower than the theoretically expected value of 9.72 $\mu_B/Tb^{3+}$ expected for $Tb^{3+}$ ions in $^7F_6$ ($4f^8$, $\mu_{eff} = g\sqrt{J(J+1)}$ $\mu_B$ = 9.72 $\mu_B$ ; $g = 1.5, S = 3, L = 3, J = 6$) [20]. Fitting parameters are summarized in Table II and agree well with previously reported values [23, 24]. For 1.8 ≤ T ≤ 50 K range, susceptibility deviates from Curie-Weiss behavior due to crystal electric field (CEF) effects and short-range spin correlations [13]. At low-temperatures (1.8 – 25 K), χ(T) is fitted with the modified Curie-Weiss (CW) form, $\chi(T) = \chi_0 + C/(T - \theta_{CW})$, where $\chi_0$ is the $T$-independent contribution and $\theta_{CW}$ is the Curie-Weiss $T$; fits are shown as solid blue lines in the insets of Fig. 2(a, b). The reduced values indicate that below 25 K, susceptibility is dominated by low-lying CEF levels separated from higher excited states, restoring the full expected moment above 50 K. The $\theta_{CW}$ reflects the energy scale of magnetic exchange ($J_{nn}$) between $Tb^{3+}$ moments, with non-Kramers doublets separated by ~ 1.5 meV governing low-$T$ single-ion properties [25]. We observed crystal field mixing states due to presence of disordered structure, revealed in INS experiments (Section D). The lack of ordered CEF states mainly governs the low-$T$ physics in these frustrated compounds. Using a mean-field approach, the nearest-neighbour interactions ($J_{nn}$), can be estimated by the relation $J_{nn}= 3\Theta/zJ(J+1)$, where z = 6 is the number of nearest neighbors, J = 6 is the total angular momentum for $Tb^{3+}$, and Θ is the $\theta_{CW}$ obtained from experimental fits. These yields $J_{nn}/k_B$ are −41(3) and −28(9) mK for x = 0 and 0.5, respectively [26, 27]. The ionic radii difference between $Zr^{4+}$ and $Ti^{4+}$ modifies $J_{nn}$ of $Tb^{3+}$ ions, placing both compounds in a highly frustrated regime compared to $Tb_2Sn_2O_7$ and $Tb_2Ti_2O_7$ [26, 23]. A qualitative insight into the crystal electric field (CEF) effects can be obtained by fitting the susceptibility data to a two-level CEF model: $\chi^{-1} = 8(T - \theta_{CW})[1 + \exp(-\Delta/k_BT)]/[\mu_{eff\_0}^2 + \mu_{eff\_1}^2\exp(-\Delta/k_BT)]$, where $\Delta/k_B$ denotes the gap between the ground−state doublet and first excited CEF level, while $\mu_{eff\_0}$ and $\mu_{eff\_1}$ represent the effective magnetic moments of the ground and excited states, respectively [28]. As shown in Fig. 2 (c, d), this model agrees well with experimental data for both compounds in the range 1.8 - 350 K, yielding $\mu_{eff\_0} = 5.31$ $\mu_B$, $\mu_{eff\_1} = 6.87$ $\mu_B$, $\theta_{CW} = -5.43$ $K$ and $\Delta/k_B = 165.53$ K (~ 14.27 meV) for x = 0, and $\mu_{eff\_0} = 6.10\mu_B, \mu_{eff\_1} = 7.57$ $\mu_B$, $\theta_{CW} = -9.22$ $K$ and $\Delta/k_B = 182.87$ K (~15.76 meV) for x = 0.50. Negative CW $T$ highlights dominant AFM correlations between $Tb^{3+}$ ions. For low-energy CEF structure, we refit the data with this model over the restricted 1.8 - 25 K range [insets of Fig. 2 (c, d)], which reveals a smaller gap of $\Delta/k_B = 16.815$ K (~1.4 meV) and 19.44 K (~1.68 meV), with CW temperatures (- 2.82 K and - 2.27 K) for the compounds x = 0 and x = 0.50 respectively. Similarly, weak excitation energy modes appear near 2.0 meV in the INS experimental data at $E_i$ = 11.2 meV with T = 4 K (Fig. S 1 of Supplementary Information). Interestingly, the values of $\Delta/k_B$ values align closely with point-charge model CEF calculations using the PyCrystalField code [Supplementary Information]. Collectively, all these energy scales (INS experimental, CEF fit, PyCrystalField code calculation) indicate closely spaced lowest CEF levels, rendering them susceptible to mixing from structural instability, vacancies, and defects.

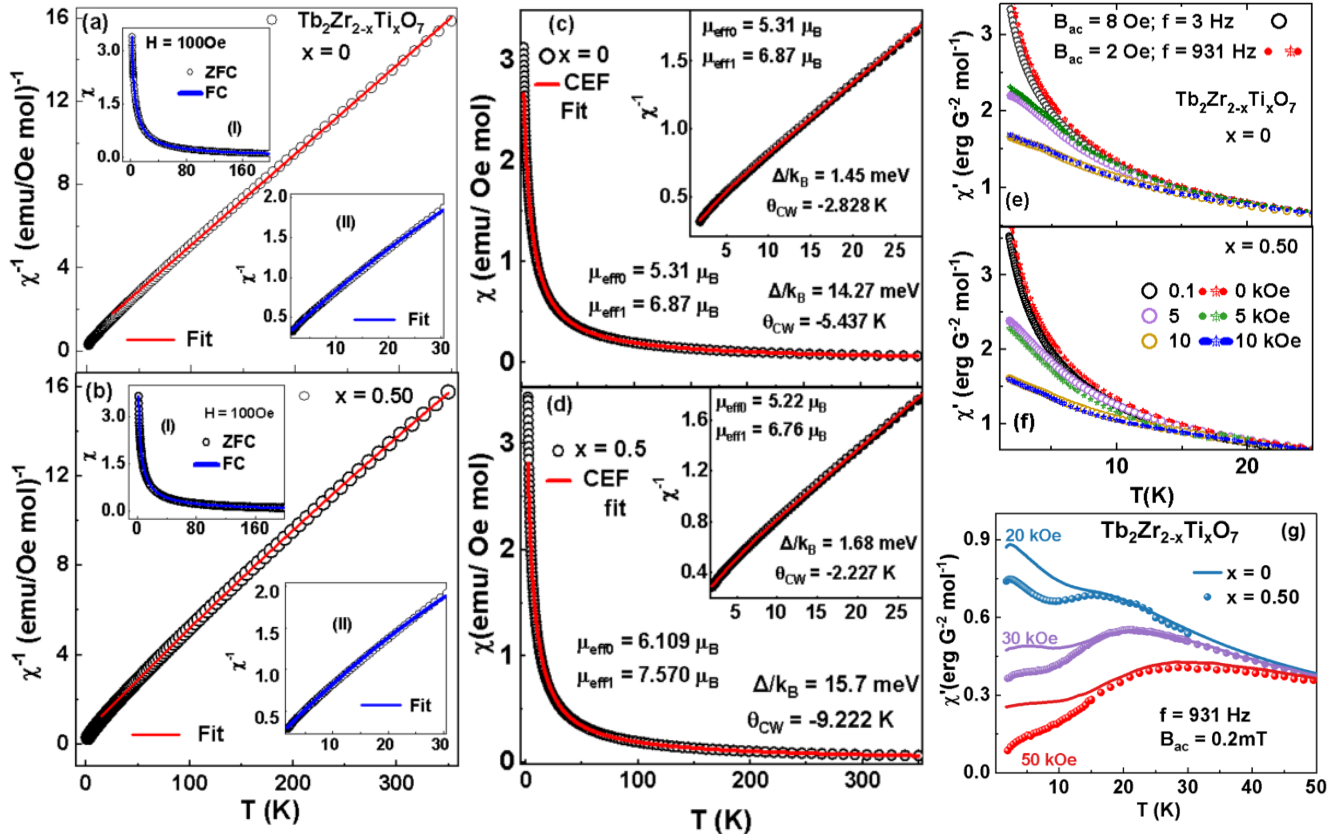

Figure 2. (a, b) DC susceptibility $\chi_{dc}$ measured at H = 100 Oe for x = 0 and x = 0.5. Solid red lines denote high-$T$ Curie-Weiss fits; insets show (i) overlapping ZFC and FC curves, and (ii) low-$T$ modified CW fits (solid blue lines). (c, d) Two-level crystal electric field fit for x = 0 and x = 0.5 in the range of 1.8 – 350 K; (inset - in the range of 1.8 -25 K). The $T$ dependence of $\chi'(T)$ measured at dc applied fields ($B_{dc}$) of 0, 0.1, 5, and 10 kOe (e) x = 0 and (f) x = 0.50, data measured at low-frequency ($f$ =3 Hz, $B_{ac}$=8 Oe) and high frequency ($f$ = 931 Hz, $B_{ac}$ = 2 Oe). (f) Comparison of $\chi'(T)$ for both compositions x = 0 and 0.5 at $f$ = 931 Hz under different applied fields ($B_{dc}$ = 20, 30, 50 kOe).

Table II. Curie–Weiss fit parameters obtained from high and low-$T$ data

| $Tb_2Zr_{2-x}Ti_xO_7$ | x = 0 | x = 0.50 |
|---|---|---|
| T = 50 - 350 K | | |
| $C_{free}$ (emu K Oe$^{-1}$mol$^{-1}$) | 11.43(7) | 11.59(2) |
| $\theta_C^{free}$ (K) | -16.50(7) | -19.95(4) |
| $\mu_{eff}^{free}$ ($\mu_B/Tb^{3+}$) | 9.54 | 9.59 |
| T = 1.8 - 25 K | | |
| $\chi_0$ | 0.106(2) | 0.114(8) |
| $C_{LT}$ (emuKOe$^{-1}$mol$^{-1}$) | 7.52(5) | 6.88(3) |
| $\theta_{CW}$ (K) | -3.47(2) | -2.43(3) |
| $\mu_{eff}^{LT}$ ($\mu_B/Tb^{3+}$) | 7.75 | 7.41 |

The real part of ac susceptibility $\chi'(T)$ for x = 0 and x = 0.50 were measured over 1.8 - 40 K using AC fields ($B_{ac}$ = 2 - 8 Oe) under various DC fields ($B_{dc}$ = 0 - 5 T). Zero-field $\chi'(T)$ for both compositions display paramagnetic behavior and remain frequency-independent across 1 Hz $\leq f \leq$ 1 kHz, indicating the absence of long-range magnetic order or phase transitions. The low-$f$ (3, 113 Hz; $B_{ac}$ = 8 Oe) under applied fields closely track the high-frequency data (331, 531, 731, 931 Hz; $B_{ac}$ = 2 Oe) without showing any frequency dependent shift. The high-frequency (931 Hz) data nearly overlap with low-frequency results. With increasing field, $\chi'(T)$ magnitude gradually decreases, signaling partial spin polarization and suppression of low-energy collective fluctuations.

At higher fields, these slope changes sharpen and evolve into a narrow low-$T$ peak below 5 K and a broad hump near T ≈ 20 K as shown in Fig. 2(g) for $f$ = 931 Hz. Both compositions exhibit qualitatively similar behavior, differing primarily in low-$T$ response amplitudes. The narrow peak (emerging above $B \geq$ 15 kOe) reflects partial polarization that stabilizes short-range correlations [23]. Above $B_{dc}$ = 20 kOe, the broad hump shifts to higher temperatures, indicating slow spin relaxation consistent with prior reports on $Tb_2Zr_2O_7$ [13]. Peak magnitude decreases and shifts to higher $T$ with field, mirroring cooperative paramagnetic behavior in $Tb_2Ti_2O_7$ where Zeeman energy surpasses thermal fluctuations to produce a susceptibility maximum at the polarization crossover [13, 29].

C. Onset of spin freezing at low-$T$

Low-$T$ magnetic measurements down to 0.4 K under applied fields of 0.1–100 kOe (Fig. 3) reveal irreversibility between zero-field-cooled (ZFC) and field-cooled (FC) susceptibility data below $T_{irr}$ ~1.25 K and ~1.05 K for x = 0 and 0.5, respectively. The ZFC curves exhibit a pronounced maximum at $T_{irr}$ while the FC curves approach saturation as represented in Fig. 3 (a, b), indicating a spin frozen state [30, 31]. This behavior is akin to the Coulomb spin liquid $Tb_2Hf_2O_7$, where spin dynamics gradually slow down below ~ 0.75 K ($T_{SG}$), accompanied by a bifurcation between ZFC and FC susceptibility curves, marking the onset of microscopic spin freezing on slow time scales which coexists with fast fluctuations [29, 32]. The irreversibility and glassy magnetization features are also observed in $Tb_2Ti_2O_7$, below 200 mK [31]. The clear splitting pattern are presented by subtracting ZFC and FC curves ($\chi_{FC} - \chi_{ZFC}$) in Fig. 3 (b, d). The persistent ZFC-FC bifurcation across $Tb_2M_2O_7$ (Zr, Ti and Hf) compounds indicates a common dynamic spin-freezing mechanism for Tb moments. Comparable irreversibility temperatures with varying M-site ions, which create distortions and alter $J_{nn}$, suggest that $T_{irr}$ shows weak dependence on $J_{nn}$ [27, 33]. Titanium doping introduces defects that modify the frozen magnetic degrees of freedom, yielding slightly suppressed $T_{irr}$ [32]. External field ($B_{dc} \geq$ 50 kOe) suppresses $T_{irr}$ in both compounds, as the field provides coherent alignment energy to the spins trapped in random anisotropy wells, and thus eliminating the irreversibility.

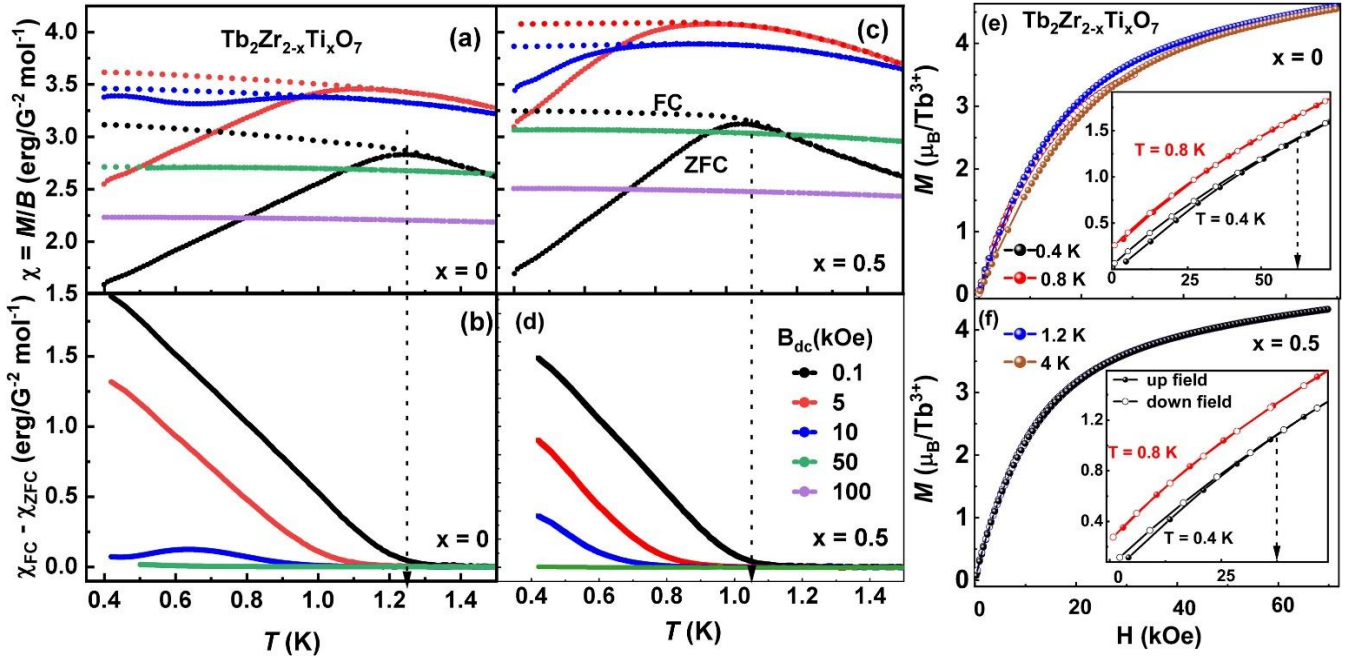

Figure 3. (a, b) Field-dependent DC susceptibility for x = 0 and x = 0.50; closed and open symbols denote field-cooled (FC) and zero-field-cooled (ZFC) data, respectively. (c, d) Field-dependent variation of $\Delta\chi = \chi_{FC} - \chi_{ZFC}$ for both compositions. (e, f) Isothermal magnetization M(H) curves at various temperatures for x = 0 and x = 0.50. Insets highlight small hysteresis loops at 0.4 K and 8 K; dashed lines mark fields where field-up (filled circles) and field-down (open circles) magnetization completely reverses. The 0.8 K data in the insets of (e, f) are vertical shift by 0.25 $\mu_B/Tb^{3+}$ for visibility.

The irreversibility in the low-$T$ spin state was further supported by isothermal magnetization M(H) curves, measured at different $T$ 0.4, 0.8, 1.2, 4, and 10 K, as shown in Fig. 3(e, f). For x = 0, at T = 0.4 K, curve exhibits a saturating trend (M$_{s\ (at\ 30\ kOe)}$ = 4.50 $\mu_B/Tb^{+3}$) but the full saturation is not attained even at 70 kOe, as seen at 1.2 K, where M$_{S\ (at\ 70\ kOe)}$ = 4.59 $\mu_B/Tb^{+3}$. For x = 0.50 compound exhibits comparable behavior at T = 0.4 K, with M$_{S\ (at\ 70\ kOe)}$ = 4.35$\mu_B/Tb^{3+}$ consistent with x = 0, and other previous study on $Tb^{3+}$ [24]. The M$_S$ value is significantly smaller than the theoretically expected M$_S \cong$ 9 $\mu_B$ (for g =1.5 and J = 6) of free $Tb^{3+}$ ions, recovering only ~ 50% of the full value. This reduction signals robust crystal-field anisotropy and persistent short-range correlations that prevent complete moment alignment [13, 29]. The spin freezing manifests as small hysteresis in M(H) below 1.8 K, measurements from 0–70 kOe (field up) and 70 – 0 kOe (field down) reveals clear hysteresis down to 0.4 K in both compounds as shown in inset of Fig. 3 (e, f). The coercive gap between the field-up and field-down curve measures ~320 Oe for x = 0 and is ~200 Oe for x = 0.50 at 0.4 K. As expected, M(H) becomes reversible again above 1.2 K with hysteresis loops fully closing by 1.8 K.

D. Inelastic Neutron Scattering

We performed INS measurements at 4 and 100 K with various neutron energies, to understand the CEF energy levels. However, as shown in Fig. 4 for 4 K, both compounds exhibit broad diffuse magnetic scattering (solid red box) at low-momentum transfer Q, indicating that lattice disorder strongly affects the single-ion magnetic environment and suppresses well-defined CEF excitations. At high Q, the marked dashed box region, also exhibits broad and dispersive features originating from lattice vibrations and partial overlapping of the magnetic signals. We have shown the dynamic structure factor $S(Q, E)$, in Fig. 5, obtained by summing the scattering intensities over both low- and high-Q regions at 4 and 100 K for various incident neutron energies ($E_i$). Fig. 5(a) demonstrates that the momentum-transfer dependence of $S(Q, E)$ at $E_i = 27.5$ meV for $Tb_2Zr_2O_7$, follows the dipolar magnetic form factor characteristic of the $Tb^{3+}$ ion, confirming the magnetic origin of the inelastic peaks [34]. The progressive reduction in intensity with increasing Q reflects the attenuation of the neutron magnetic scattering cross-section due to the spatially extended *4f* electron density of Tb ions. This trend, evident in both compounds (Fig. 5 (a, c, d, e)), indicates that the observed excitations originate from transitions within the CEF levels of the $Tb^{3+}$ ground multiplet. Notably, instead of sharp CEF peaks, broad diffuse magnetic peaks appear in the 3–10 meV range. The 4 K magnetic excitations for $Tb_2Zr_{1.5}Ti_{0.5}O_7$ at $E_i = 27, 65, 150$ meV (Fig. 5 (b) and inset), are very weak and dispersive. For $E_i = 65$ meV, the spectrum is more dispersive, extending up to ~ 40 meV with noticeable Q-dependent intensity variations for both compounds (Fig. 5 (c, d)). The inset of Fig. 5 (c) shows the spectra at 4 and 100 K for $Tb_2Zr_2O_7$ at $E_i = 65$ meV, revealing reduced coherence of short-range magnetic correlations at higher *T*, which weakens the diffuse magnetic peak intensity. At 4 K for $Tb_2Zr_2O_7$, $S(Q, E)$ shows a peak near 20 meV at high Q, most likely due to phonon contribution (Fig.

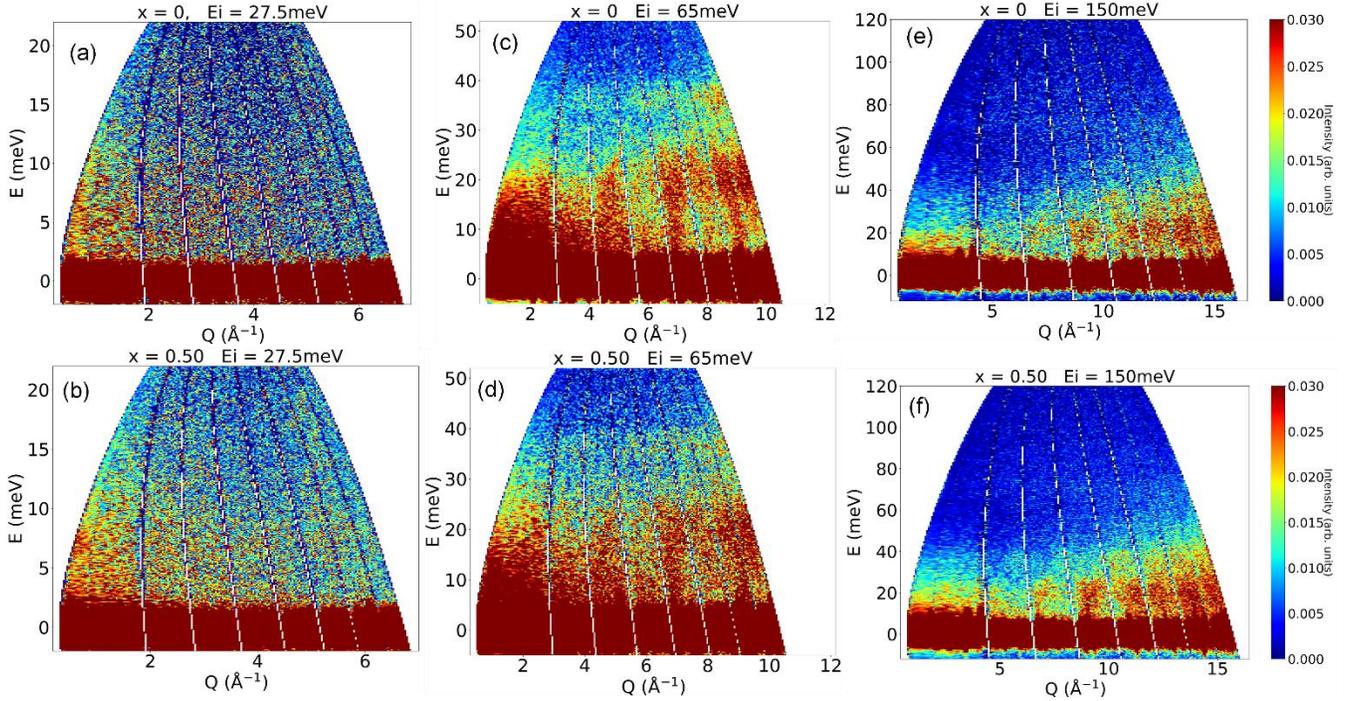

Figure 4. Contour intensity maps for $Tb_2Zr_{2-x}Ti_xO_7$ collected at 4 K on MARI with incident neutrons of (a, b) $E_i =$ 27.5 meV, (c, d) $E_i = 65$ meV, and (e, f) $E_i = 150$ meV for compositions x = 0 and x = 0.50, respectively. The red arrow indicates the CEF excitation, the red solid box marks the diffusive magnetic scattering, and the purple dashed box denotes the phonon contribution.

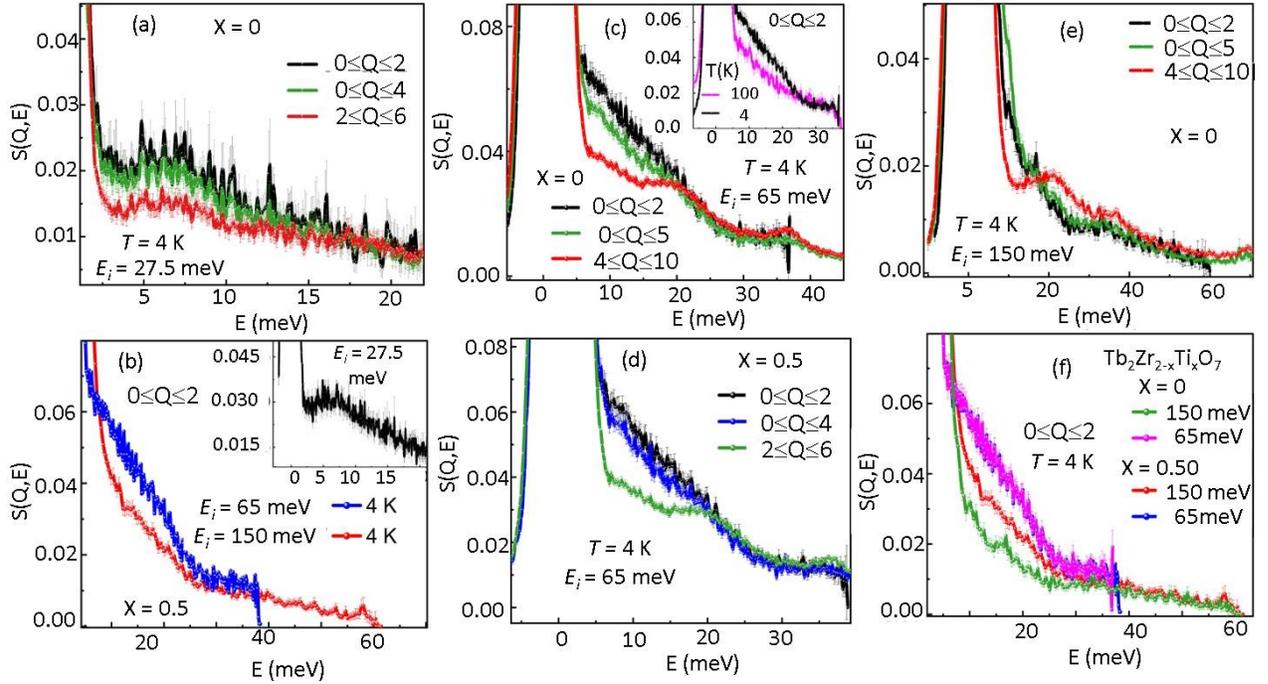

Figure 5. The INS Q-cut spectra at 4 K for $Tb_2Zr_{2-x}Ti_xO_7$ pyrochlores (x = 0, 0.50). (a) Q-dependence of the 27.5 meV excitation in x = 0. (b) Low-Q cuts for $E_i$ = 65, 150, and 27.5 meV(inset) in x = 0.50. (c) Q-dependence at $E_i$ = 65 meV for x = 0; inset shows low-Q cuts at 4 and 100 K. (d) Q-dependence at $E_i$ = 65 meV for x = 0.50. (e) Q-dependence at $E_i$ = 150 meV for x = 0. (f) Low-Q comparison at $E_i$ = 65 and 150 meV for both compounds.

5(e)). A comparison of the 4 K spectra for both compounds at $E_i$ = 65, 150 meV is presented in Fig. 5(f). Although the low-energy response is similar in both samples, the $E_i$ = 150 meV intensity is weaker in the parent compound. This suggests that, cation and anion disorder in the defect fluorite phase disrupts coherent Tb–Tb spin correlations, while the doped compound sustains stronger low-energy magnetic response and spectral weight. It is worth mentioning that for $E_i$ = 11.2 meV, a weak magnetic excitation ~ 2meV is observed at low-Q in fluorite structure. Similar low-energy modes in related rare-earth pyrochlores have been attributed to hybrid magnetoelastic excitations influenced by local strain and defects and is fitted with a Lorentzian profile (mentioned in Supplementary Information Fig S1) [34, 35]. This excitation likely arises from low-energy CEF excitations of $Tb^{3+}$ ions coupled with magnetic correlations and magnetoelastic interactions linked to the disordered structure.

Further, we used point-charge model to estimate the CEF energy levels (shown in Supplementary Information) for both compounds. However, it was difficult to match these with obtained diffuse spectrum in our INS measurements. These broad features result from overlapping CEF levels enhanced by collective effects such as spin fluctuations or quasi-elastic scattering due to dynamic disorder. This behavior reflects quenched random crystal fields caused by a high density of anion defects, which broaden and suppress sharp CEF modes, as detailed by Sibille et al. [32]. In another Tb based pyrochlore $Tb_2ScNbO_7$, similar diffuse INS intensity is attributed to charge disorder, leading to suppressed CEF and magnon modes and a collective magnetic state dominated by broad continua which was attributed to structural disorder and magnetoelastic coupling [11]. Thus, these observations suggest a combined influence of structural disorder and crystal field interactions on $Tb^{3+}$ ions, which is consistent with previous neutron scattering studies on rare-earth pyrochlores.

**DISCUSSION**

$Tb_2Zr_2O_7$ intrinsically exhibits cation and anion disorder owing to larger $Zr$ cation size and its chemical preferences relative to $Ti$. This structural instability promotes defect-fluorite phases, and influences magnetic frustration and spin dynamics as observed in $Tb_2Zr_2O_7$ and $Ho_2Zr_2O_7$ [13, 36]. Similar structural phase change occurs in $Ho_2Ti_{2-x}Zr_xO_7$, where substitution at M site induces changes in symmetry [20, 37]. $Tb_2Zr_2O_7$ and $Tb_2Zr_{1.50}Ti_{0.50}O_7$ exhibit a characteristic frequency-independent peak in $\chi'(T)$, unlike the spin-glass-like behavior reported in a prior $Tb_2Zr_2O_7$ study [13]. These contrasting dynamics likely arise from variations in sample homogeneity, grain size, and defect concentrations induced by differing synthesis methods.

The emergence of spin-freezing phenomena across $R_2M_2O_7$ (R = Dy, Ho, Tb; M = Ti, Sn, Zr, Hf) pyrochlore family reveals a complex interplay between geometric frustration, lattice disorder, and slow spin dynamics. In classical spin ice systems $Ho_2Ti_2O_7$ and $Dy_2Ti_2O_7$, the freezing is reported at $T_{irr} \cong 0.6$ - 0.75 K and largely independent of the $J_{nn}$, suggesting that the onset of slow dynamics is a universal feature intrinsic to the pyrochlore lattice [29, 38, 39]. In $Dy_2Ti_2O_7$, where M(H) loops exhibit coercive gaps below ~ 0.65 K and become fully reversible at ~ 0.8 K. In this spin ice, spins are highly uniaxial, unlike the isotropic spins in geometrically frustrated antiferromagnets showing spin freezing; these differences imply disorder or magnetic lattice dilution drives the observed freezing [38]. In the present study, the spin-freezing of both compositions is evidenced by ZFC-FC irreversibility ($T_{irr}$ = 1.25 K, 1.05 K, respectively) along with the narrow magnetic hysteresis loops at 0.4 K. The $T_{irr}$ shift systematically with $Ti$-doping, indicating modified freezing driven by enhanced structural disorder and changes in $J_{nn}$ exchange pathways. The hysteresis loops are rapidly suppressed on increasing $T$. This sensitivity to $T$ and external fields, along with the doping-dependent evolution of the magnetic response, distinguishes the low-$T$ defect-driven correlated freezing regime characterized by short-range spin correlations and slow spin dynamics in x = 0 and x = 0.50 compounds. Weak magnetic hysteresis was reported in $Tb_2Ti_2O_7$, where this history dependence arises from extremely slow spin dynamics, is suppressed by modest applied fields (~0.36 T) [40]. Collectively, the field-sensitivity and relaxation scales observed here suggest that Tb-based pyrochlores occupy a unique regime between classical freezing and quantum fluctuations.

The INS spectra of both compounds are characterized by weak low-energy excitations and a broad diffuse magnetic continuum, indicating that structural disorder or vacancies significantly smear the magnetic response and suppress the formation of well-defined excitation modes. In contrast, $Tb_2Ti_2O_7$ exhibits sharp crystal electric field (CEF) excitations along with magnetoelastic hybrid modes and persistent spin-liquid-like dynamics [25, 35, 41]. Comparison highlights, how disorder disrupts coherent spin excitations in the present system, yielding a correlated, diffuse magnetic response. A similar disorder-driven magnetic response has been reported in $Tb_2ScNbO_7$, where charge and site mixing on the nonmagnetic sublattice lead to significant broadening and overlap of crystal electric field (CEF) excitations. As a result, the magnetic spectra are dominated by diffuse inelastic continua, reflecting the presence of quenched random fields and enhanced low-energy spin fluctuations [11], where a high density of anion Frenkel defects strongly distorts the local crystal field environment, further suppressing well-defined excitations and promoting a continuum-like magnetic response [32]. INS spectra reveal broadened low-energy CEF excitations and diffuse magnetic scattering, consistent with a Coulomb spin-liquid–like regime that evolves into glassy behavior at low-temperatures. Taken together, these observations underscore the pivotal role of disorder in destabilizing coherent excitations and driving the system toward a dynamically fluctuating yet spatially correlated magnetic state [32]. A similar scenario is realized in defect-fluorite $Ho_2Zr_2O_7$, where the nonmagnetic crystal electric field (CEF) ground state is not well separated from the excited levels, in contrast to the well-isolated Ising doublet of $Ho_2Ti_2O_7$ [36]. These unifying behavior highlights how local disorder perturbs the CEF scheme and weakens the definition of magnetic excitations, even in the absence of long-range structural disruption. In aggregate, these results establish a consistent picture of disorder-driven modification of CEF states across rare-earth pyrochlore and fluorite materials, where similar smearing of excitations emerges despite differing microscopic origins of disorder. Across the $Tb_2B_2O_7$ series, increasing structural disorder gradually lowers local symmetry, transforming sharp CEF excitations in ordered pyrochlores into broad, Q-dependent magnetic continua in disordered compounds. Notably, even partial disorder is sufficient to blur the CEF spectrum and diminish coherent excitations, underscoring the central role of anion and cation disorder in governing low-energy spin dynamics.

## CONCLUSIONS

Our results provide evidence for a macroscopic spin-freezing state below ~ 1.25 K and ~ 1.05 K for $Tb_2Zr_2O_7$ and $Tb_2Zr_{1.5}Ti_{0.5}O_7$, respectively. This freezing is manifested through clear ZFC–FC irreversibility together with narrow magnetic hysteresis loops observed at 0.4 K with coercive gaps (~ 320 Oe and ~ 200 Oe, respectively) between field up and field down. The hysteresis loops are confined at $H$ = 6 kOe for x = 0 and $H$ = 3 kOe for x = 0.50 and are rapidly suppressed with increasing $T$, indicating a fragile frozen state that is highly sensitive to both thermal fluctuations and external magnetic fields. The shift in the spin-freezing $T$ with Ti substitution further indicates a modification of the spin dynamics, nearest neighbour exchange parameters and changes in the structure and magnetic environment. At higher $T$ (~ 20 K), a persistent slow spin dynamics shown under applied fields above 20 kOe, points towards a broad regime of low-energy correlated spin fluctuations beyond the freezing $T$. Inelastic neutron scattering reveals that very weak and unresolved excitations are visible at $E_i$ = 11.2 meV, whereas at higher incident energies ($E_i$ = 27.5 meV, 65 meV, and 150 meV), the results represent stronger broadening and a diffuse, continuum-like character. This reflects the combined effects of structural disorder, spin freezing effects and short-range magnetic correlations, which gradually broaden the crystal electric field excitations. These results support a correlated, disorder-influenced magnetic state in Tb-based disordered fluorites, with modifications introduced with Ti substitution at the Zr site.

## ACKNOWLEDGEMENT

S.S. acknowledges the Ministry of Education (MoE), India, for the senior research fellowship (HTRA). We thank the Advanced Materials Research Center (AMRC), IIT Mandi, for experimental facilities. Neutron scattering experiments at the ISIS Neutron and Muon Source were supported by beamtime allocation RB2520154 from the Science and Technology Facilities Council (STFC).

# Supplementary Information

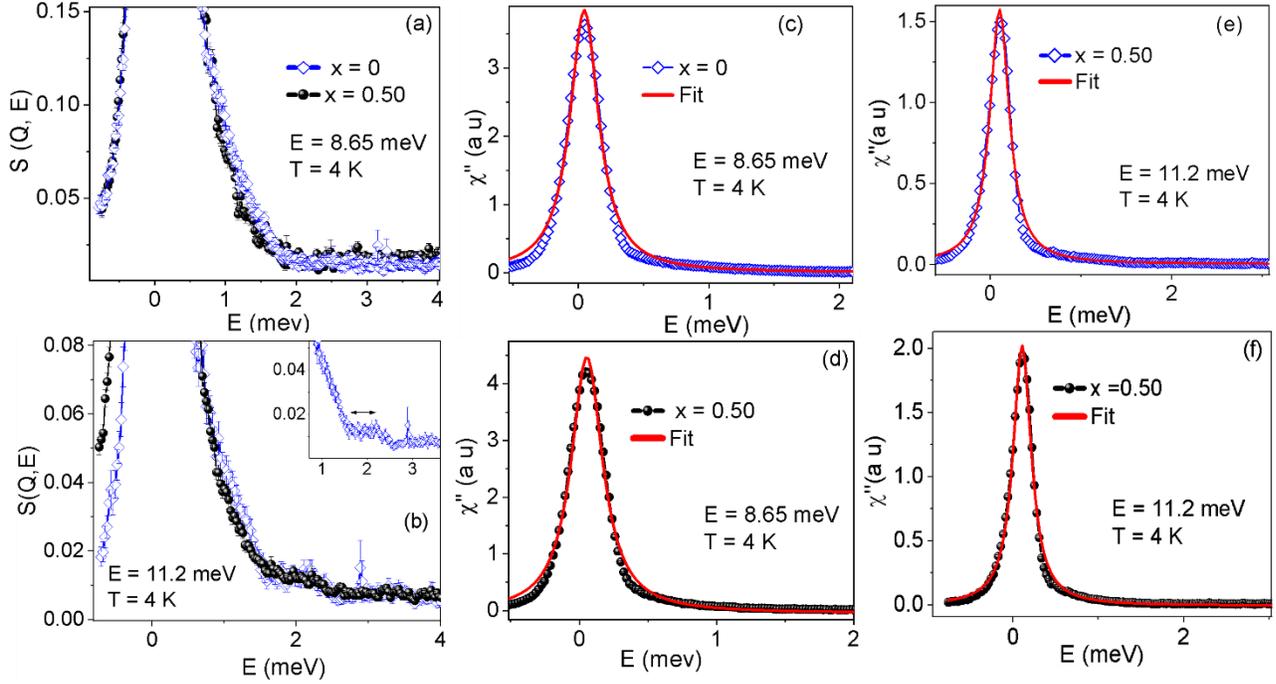

Figure S1. The Low energy inelastic neutron scattering spectra measured at 4 K in the elastic momentum transfer range Q ≤ 2 Å$^{-1}$. The data show a broad continuum-like response, which may be related to spin freezing in the system. The dynamical structure factor S (Q, E) with error bars is presented for incident energies (a) $E_i$ = 8.65 meV and (b) $E_i$ =11.2 meV. Data for x=0 (blue) and x=0.50 (black) are shown. (b) - inset: highlights a weak excitation near ~ 2 meV. Its intensity is significant smaller relative to the nearest prominent peak; therefore, it is not fitted as a separate peak and is only indicated as a weak mode. The imaginary part of the dynamic spin susceptibility is calculated using $\chi''(E) = S(E) \times (1 - \exp(-E/k_B T))$ where E is the neutron energy transfer and $k_B$ are Boltzmann's constant. We fit (red line) the data of $\chi''(E) = \Gamma E/((E - \Delta)^2 + \Gamma^2)$ where Γ represents the damping (linewidth) and Δ is the characteristic energy scale. $E_i$ = 8.65 meV, the fitted parameters are $E_i$ = 8.65 meV: (c) x = 0 compound are Γ = 0.2690 ± 0.0046 meV and Δ = 0.0456 ± 0.0014 meV, (d) x = 0.50 compound are Γ = 0.2788 ± 0.0046 meV and Δ = 0.0550 ± 0.0014 meV. For $E_i$ = 11.2 meV, the fitted parameters are: (e) x = 0 compound are Γ = 0.2621 ± 0.0042 meV and Δ = 0.2621 ± 0.0140 meV, (f) x = 0.50 compound are Γ = 0.2576 ± 0.0030 meV and Δ = 0.1070 ± 0.0014 meV.

The lowest-lying crystal electric field (CEF) excitations in $Tb_2Ti_2O_7$ and $Tb_2Sn_2O_7$ predominantly probe transitions out of the ground-state doublet at T=1.5 K, consistent with earlier reports [1, 2]. In this context, the excitation observed in Fig. S1(b) is likely associated with such low-energy CEF levels. Point-charge calculations (mentioned below for $Tb_2Zr_2O_7$) indicate an excited state near ~1.7 meV, suggesting that the observed broad feature around ~2 meV corresponds to a smeared, plateau-like response rather than a well-defined mode. So, excitation observed in above Fig. S1 (b) is might be connected with low lying energy level. The linewidth Γ, extracted from fit the spectra in Fig. S1 (c-f) extracted from fits to the spectra, reflects a combination of crystal field disorder and magnetoelastic coupling, and may also be influenced by multipolar fluctuations. Although the continuum-like character of the spectra may be suggestive of fractionalized excitations, further theoretical analysis and other experiments are required to substantiate this interpretation.

**POINT CHARGE MODEL:**

The crystal-field parameters in $Tb_2Zr_2O_7$ and $Tb_2Zr_{1.5}Ti_{0.5}O_7$ were calculated using the PyCrystalField code [3], based on crystallographic information derived from XRD - CIF files. This method treats each ligand as a point charge and calculates the electrostatic potential at the $Tb^{3+}$ site by applying the local coordination geometry and symmetry axes. PyCrystalField employs group-theoretical constraints to determine the crystal-field parameters and subsequently computes the crystal-field energy levels and eigenvector compositions for the 4f electron manifold of $Tb^{3+}$ ($J = 6$; $2J + 1 = 13$). The crystal-field Hamiltonian, expressed as $H_{CEF} = \sum_{i,j} B_m^n O_m^n$; where $B_m^n$ denotes the crystal field parameters and $O_m^n$ are the Steven's operators [4]. This model is highly sensitive to the structural parameters, and even minor variations in atomic positions and vacancies can lead to significant shifts in the CEF energies. However, XRD reveals distinct crystal structures for the both compounds $Tb_2Zr_2O_7$ and $Tb_2Zr_{1.5}Ti_{0.5}O_7$, reflected in their CEF calculations, which show variations in energy level schemes. For $Tb_2Zr_2O_7$, CEF parameters were calculated by considering the structural models as defect fluorite phase (space group $Fm-3m$, No. 225) and the doped compound $Tb_2Zr_{1.5}Ti_{0.5}O_7$ as pyrochlore phase ($Fd-3m$, No. 227). The CEF energy levels and wavefunctions, obtained by diagonalizing the CEF Hamiltonian matrices, are presented below. These levels are difficult to extract directly from inelastic neutron scattering data due to the broad, diffuse scattering behavior and the apparent absence of distinct CEF excitations.

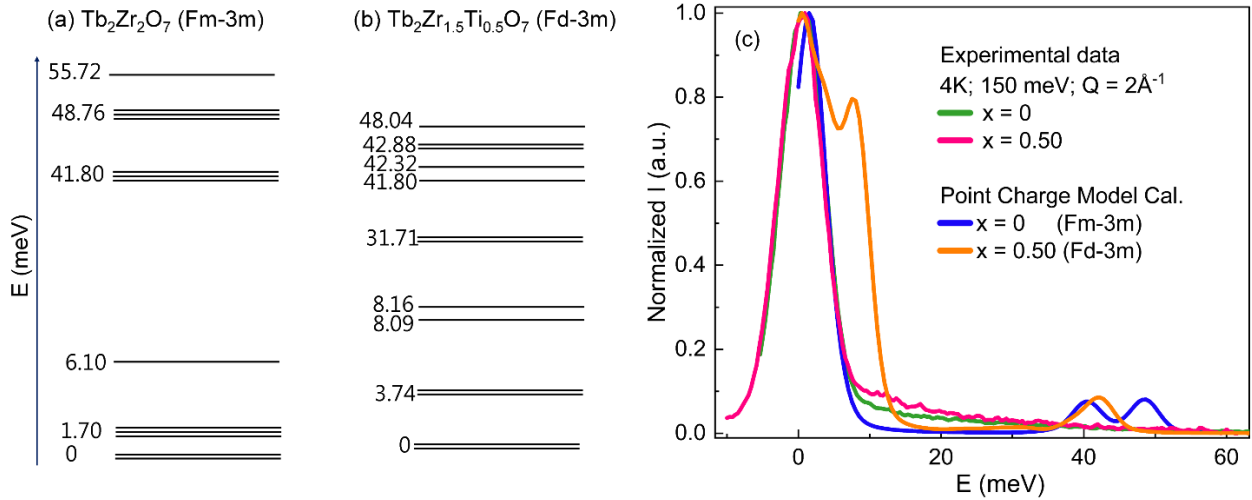

Figure S2. CEF level schemes for all 2J+1=13 states of $Tb^{3+}$ ions in (a) $Tb_2Zr_2O_7$ ($Fm-3m$) and (b) $Tb_2Zr_{1.5}Ti_{0.5}O_7$ ($Fd-3m$), as determined from point-charge model calculations. These diagrams highlight the influence of structural symmetry lowering from ordered pyrochlore to disordered fluorite as well as the impact of B-site chemical substitution on the splitting and arrangement of the $Tb^{3+}$ crystal field multiplet. The corresponding wavefunctions are provided below for reference. (c) Comparison of experimental inelastic neutron scattering spectra (4 K, 150 meV incident energy, Q = 2 Å$^{-1}$) for $Tb_2Zr_{2-x}Ti_xO_7$ with x = 0 and x = 0.50 compositions (solid green and magenta lines, respectively) alongside calculated spectra using the Point Charge Model using both pyrochlore ($Fd-3m$ orange lines) and disordered fluorite ($Fm-3m$, blue line) crystal structures. The intensity is normalized to the maximum peak for direct comparison.

*Tb₂Zr₂O₇* (Space group 225)

$|\psi_0^0\rangle = \pm 0.586 |\pm 6\rangle + 0.250 |\pm 4\rangle \pm 0.395 |\pm 2\rangle - 0.935 |0\rangle$

$|\psi_1^{1.70}\rangle = -0.705 |\pm 6\rangle - 0.165 |\pm 5\rangle - 0.423 |\pm 3\rangle + 0.050 |\pm 2\rangle + 0.891 |\pm 1\rangle$

$|\psi_2^{6.10}\rangle = \pm 0.395 |\pm 6\rangle \pm 0.586 |\pm 2\rangle$

$|\psi_3^{41.80}\rangle = -0.050 |\pm 6\rangle \pm 0.793 |\pm 5\rangle \pm 0.594 |\pm 3\rangle - 0.705 |\pm 2\rangle \pm 0.135 |\pm 1\rangle$

$|\psi_4^{48.76}\rangle = \pm 0.586 |\pm 5\rangle \pm 0.707 |\pm 4\rangle \pm 0.685 |\pm 3\rangle \pm 0.433 |\pm 1\rangle$

$|\psi_5^{55.72}\rangle = -0.661 |\pm 4\rangle - 0.354 |0\rangle$

Tb₂Zr₁.₅Ti₀.₅₀O₇ (Space group 227)

$|\psi_0^0\rangle = \pm 0.003 |\pm 5\rangle + 0.973 |\pm 4\rangle + 0.019 |\pm 2\rangle \pm 0.232 |\pm 1\rangle$

$|\psi_1^{3.74}\rangle = 0.906 |\pm 5\rangle \pm 0.42 |\pm 2\rangle + 0.045 |\pm 1\rangle$

$|\psi_2^{8.09}\rangle = -0.184 |\pm 6\rangle \pm 0.662 |\pm 3\rangle - 0.236 |0\rangle$

$|\psi_3^{8.16}\rangle = \pm 0.186 |\pm 6\rangle - 0.682 |\pm 3\rangle$

$|\psi_4^{31.71}\rangle = 0.421 |\pm 5\rangle \pm 0.025 |\pm 4\rangle \pm 0.89 |\pm 2\rangle - 0.173 |\pm 1\rangle$

$|\psi_5^{41.80}\rangle = 0.658 |\pm 6\rangle \pm 0.125 |\pm 3\rangle - 0.321 |0\rangle$

$|\psi_6^{42.32}\rangle = \pm 0.682 |\pm 6\rangle - 0.186 |\pm 3\rangle$

$|\psi_7^{42.88}\rangle = 0.034 |\pm 5\rangle + 0.231 |\pm 4\rangle - 0.176 |\pm 2\rangle \pm 0.956 |\pm 1\rangle$

$|\psi_8^{48.04}\rangle = -0.183 |\pm 6\rangle \pm 0.214 |\pm 3\rangle - 0.917 |0\rangle$